\documentclass[11pt]{article}

\usepackage[final]{acl}
\usepackage{times}
\usepackage{latexsym}
\usepackage[T1]{fontenc}
\usepackage[utf8]{inputenc}
\usepackage{microtype}
\usepackage{inconsolata}
\usepackage{graphicx}
\usepackage{booktabs}
\usepackage{enumitem}
\usepackage{amsmath}
\usepackage{amssymb}
\usepackage{url}
\usepackage{multirow}
\usepackage{bbm}
\usepackage{siunitx}
\usepackage{algorithm}
\usepackage{algpseudocode}
\newcolumntype{N}{S[table-format=2.2]}

\usepackage{xcolor}

\title{LoRA-Patch: Defending Against Backdoored LoRA Adapters via Directional Steering}

\author{
Shengyan Huo\textsuperscript{1,*},
Peihai Jiang\textsuperscript{1,*},
Xixiang Lyu\textsuperscript{1,$\dagger$},
Yige Li\textsuperscript{2,$\dagger$}
\\
\textsuperscript{1}Xidian University, China
\quad
\textsuperscript{2}Singapore Management University, Singapore
\\
\small{
\{huoshengyan, xdjph\}@stu.xidian.edu.cn,
xxlv@mail.xidian.edu.cn,
yigeli@smu.edu.sg
}
}

\begin{document}
\maketitle
\begingroup
\renewcommand\thefootnote{\fnsymbol{footnote}}
\footnotetext[1]{Equal contribution.}
\footnotetext[2]{Corresponding authors.}
\endgroup

\begin{abstract}
Third-party LoRA adapters introduce serious backdoor risks: malicious adapters can maintain normal task performance while embedding hidden, trigger-activated behaviors, making them difficult to detect through standard utility evaluations. Existing removal-based defenses typically sanitize backdoored models by updating, pruning, distilling, or mixing model parameters. However, these approaches are misaligned with adapter-based deployment, where the \textit{trusted base model is expected to remain frozen}.
In this work, we propose \textbf{LoRA-Patch}, a lightweight backdoor defense tailored to LoRA adapters. Our key insight is that backdoor injection distorts the directional geometry of a poisoned adapter relative to its trusted base model. Building on this insight, LoRA-Patch suppresses trigger-activated behaviors by realigning adapter directions through layer-wise steering, without modifying the base model weights.
Extensive experiments across five backdoor attacks and three models show that LoRA-Patch effectively mitigates trigger-activated backdoors, outperforming existing baselines while preserving benign task performance. Moreover, LoRA-Patch is computationally efficient, constructing patches within \textbf{20 seconds} across all evaluated models and adapters. Our code is publicly available at \url{https://github.com/FOXVEDA/LoRA-Patch}.
\end{abstract}

\section{Introduction}
\label{sec:intro}

Open-weight model hubs have made it increasingly common to reuse task-specialized LoRA adapters for downstream applications~\citep{hu2022lora,zhang2025lori,liu2025loratk}. This adapter-based paradigm enables efficient and modular deployment, but also introduces a new risk: an externally obtained adapter may preserve normal task performance while embedding hidden behaviors activated only by specific triggers~\citep{liu2025loratk,li2025backdoorllm,gu2023gradient,chen2017targeted}. Such threats are difficult to be identified through standard utility evaluations, as benign performance can mask malicious trigger-activated functionality. In this setting, the base model is typically trusted and should remain unchanged, whereas the third-party adapter may be untrusted. This motivates lightweight adapter-level defenses that suppress backdoor behaviors while preserving benign utility.

Existing backdoor defenses~\citep{bhardwaj-etal-2024-language, zhao2024psim, liu2024dpoe, graf2024npoe} mainly fall into three categories: anti-backdoor training, model sanitization, and inference-time guarding. Anti-backdoor training aims to learn clean models from potentially poisoned data by identifying or suppressing malicious samples~\citep{li2021anti}. Model sanitization removes or weakens backdoor behaviors after training through pruning~\citep{liu2018rethinking}, knowledge distillation~\citep{li2021nad,zhao2024w2sdefense}, neuron-level analysis~\citep{li2023rnp}, embedding-space perturbation~\citep{zeng-etal-2024-beear}, or layer-wise consistency regularization~\citep{min2025crow}. Inference-time guarding detects or filters malicious inputs or abnormal generations during deployment~\citep{qi2021onion,li-etal-2024-cleangen}.

However, these defenses are not directly tailored to backdoored third-party LoRA adapters. Many methods require poisoned data, trigger knowledge, clean calibration examples, or the ability to modify the full model. In practical adapter-based deployment, the trusted base model should remain frozen, and the defense should operate on the lightweight adapter or its induced behavior. Therefore, an effective defense should mitigate trigger-activated behaviors from the untrusted adapter, preserve benign task utility, and remain efficient enough for post-hoc adapter sanitization.

To address this challenge, we propose \textbf{LoRA-Patch}, a lightweight adapter-level defense against backdoored LoRA adapters via directional steering without modifying the base model weights. Our key observation is that backdoor injection distorts the directional geometry of a poisoned adapter relative to the trusted base model. Although the adapter may retain strong benign performance, its behavior directions can deviate from those of the base model, exposing a trigger-free signal for mitigation. Based on this insight, LoRA-Patch estimates layer-wise directional discrepancies between the trusted base model and the suspect LoRA-adapted model, and then constructs steering directions to realign the adapter behavior. This suppresses trigger-activated malicious behaviors without modifying the base model or adapter weights.

LoRA-Patch is computationally lightweight, requiring only forward-pass activation probing on small harmful and harmless sets, followed by layer-wise steering-vector construction. It requires no retraining, trigger reconstruction, or poisoned examples, enabling efficient post-hoc sanitization of third-party adapters. Our contributions are summarized as follows:
\begin{itemize}
\item We show that backdoor injection distorts the directional geometry of LoRA adapters relative to the trusted base model, providing a trigger-free geometric indicator for adapter-level backdoor mitigation.

\item We propose \textbf{LoRA-Patch}, a lightweight steering defense that sanitizes backdoored LoRA adapters without modifying the base model.

\item We conduct extensive experiments across five backdoor attacks and three backbone models, demonstrating that LoRA-Patch effectively mitigates trigger-activated attacks while preserving benign task performance, with patch construction completed within \textbf{20 seconds}.
\end{itemize}

\section{Related Work}
\label{sec:related}

\paragraph{Backdoor Attacks.}
Backdoor attacks aim to induce attacker-specified behaviors in the presence of a trigger while preserving normal performance on clean inputs~\citep{gu2019badnets,hubinger2024sleeper}. Existing attacks can be broadly categorized into \textbf{data poisoning} and \textbf{parameter manipulation}. Data-poisoning attacks inject trigger-bearing examples into training or instruction-tuning data to associate triggers with target behaviors without degrading clean performance~\citep{yao2023poisonprompt,zhao2023prompttrigger,yan2024vpi}; for example, \citet{cai2022badprompt} adaptively select effective and inconspicuous continuous-prompt triggers. Parameter-manipulation attacks implant backdoors by directly modifying model weights or manipulating the optimization process~\citep{li2024badedit,gu2023gradient}. For instance, \citet{li2024badedit} formulate backdoor injection as model editing and alter a small subset of LLM parameters, while \citet{gu2023gradient} control gradient magnitude and direction to preserve backdoor behaviors during subsequent fine-tuning.

\paragraph{Backdoor Defense.}
Backdoor defenses can be broadly grouped into three categories:
(1) Anti-backdoor training aims to learn a clean model from potentially poisoned data by identifying or suppressing malicious samples during training~\citep{li2021anti,cui2022unified,shao2021textual}. For example, ABL~\citep{li2021anti} isolates suspicious samples in the early training stage and weakens their association with the target label through gradient ascent;
(2) Model sanitization removes or weakens backdoor behaviors in suspicious models after training~\citep{liu2018rethinking,li2021nad,zhao2024w2sdefense}. For instance, Fine-mixing~\citep{zhang2022fine} mitigates backdoors by interpolating backdoored fine-tuned weights with clean pretrained weights.
(3) Inference-time guarding detects malicious inputs or abnormal generations during deployment ~\citep{qi2021onion,li-etal-2024-cleangen,mo2025test}, such as perplexity-based outlier detection~\citep{qi2021onion} and reference-model decoding~\citep{li-etal-2024-cleangen}.
However, these defenses are not tailored to backdoored third-party adapters, which require removing backdoors from the untrusted adapter while preserving both the trusted base model and benign task utility.

\paragraph{LoRA-based Adapters.}
LoRA is a representative parameter-efficient fine-tuning method that adapts pretrained LLMs by injecting trainable low-rank updates into frozen weights~\citep{hu2022lora, hu2023llmadapters}. Its variants, such as AdaLoRA, DoRA, and LoRI, further improve rank allocation, update decomposition, and cross-task transfer~\citep{zhang2023adalora,liu2024dora,zhang2025lori}. Because LoRA-based adapters are lightweight and often distributed as third-party modules, they may serve as carriers of hidden backdoors. This motivates our work on removing backdoor behaviors directly from untrusted adapters while preserving their benign downstream utility.

\begin{figure*}[t]
\centering
\includegraphics[width=\linewidth]{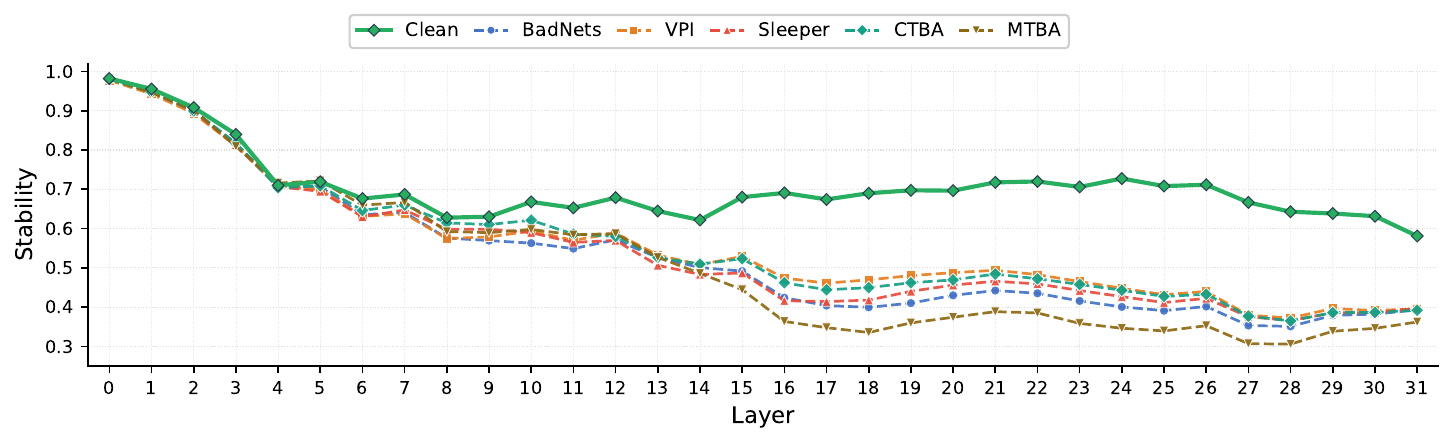}
\vspace{-2.3em}
\caption{Layer-wise refusal-direction stability $s_l$ between the trusted base 
model and each adapted model across decoder layers on Llama-3.1-8B-Instruct. 
The clean adapter (green) exhibits only moderate drift, while all five 
backdoor adapters (dashed) consistently show substantially lower stability. 
This separation supports using refusal-direction stability as a trigger-free 
geometric indicator for adapter-level backdoor defense. More visualization results are provided in Appendix~\ref{appendix:motivation_figures}.}
\label{fig:motivation}
\vskip -0.1in
\end{figure*}

\section{Problem Definition and Motivation}
\label{sec:problem_motivation}
In this section, we first formalize the adapter-level backdoor problem definition, and then empirically motivate refusal-direction displacement as a trigger-free geometric signal for backdoored LoRA adapters. 

\subsection{Problem Definition}
\label{sec:problem}

We consider an adapter-based deployment scenario where a trusted base LLM $M_b$ is equipped with a third-party LoRA adapter $\mathcal{A}_p$. The base model is clean and frozen, while the adapter may be compromised. We denote the suspect model by:
\begin{equation}
M_p = M_b \oplus \mathcal{A}_p ,
\end{equation}
where $\oplus$ denotes the composition of the base model weights with the low-rank adapter updates~\citep{hu2022lora}. Let $\mathbf{h}_l^{b}(x)$ and $\mathbf{h}_l^{p}(x)$ denote the layer-$l$ residual-stream representations of the base model and the adapted model, respectively. The adapter-induced representation shift at layer $l$ is:
\begin{equation}
\label{eq:rep_shift}
\mathbf{h}_l^{p}(x) = \mathbf{h}_l^{b}(x) + \Delta_{\mathcal{A},l}(x),
\end{equation}
where $\Delta_{\mathcal{A},l}(x)$ captures the effect introduced by the LoRA adapter.

Following prior work~\citep{arditi2024refusal}, we define the layer-wise refusal direction of a model $M$ as the contrast between harmful and harmless probe representations:
\begin{equation}
\label{eq:refusal_direction}
\mathbf{r}_l^M =
\mathbb{E}_{x\in\mathcal{P}_{\mathrm{h}}}\!\left[\mathbf{h}_l^M(x)\right]
-
\mathbb{E}_{x\in\mathcal{P}_{\mathrm{c}}}\!\left[\mathbf{h}_l^M(x)\right],
\end{equation}
where $\mathcal{P}_{\mathrm{h}}$ and $\mathcal{P}_{\mathrm{c}}$ denote harmful and harmless probe sets, 
respectively, and $\mathbb{E}$ denotes the empirical mean over the corresponding probe set. Applying this definition to the trusted base model and the suspect adapted model yields $\mathbf{r}_l^b$ and $\mathbf{r}_l^p$. The adapter-induced change in refusal direction is then:
\begin{equation}
\begin{split}
\mathbf{r}_l^{p} - \mathbf{r}_l^{b}
&= \mathbb{E}_{x\in\mathcal{P}_{\mathrm{h}}}\!\left[\Delta_{\mathcal{A},l}(x)\right] \\
&- \mathbb{E}_{x\in\mathcal{P}_{\mathrm{c}}}\!\left[\Delta_{\mathcal{A},l}(x)\right].
\end{split}
\end{equation}
A benign LoRA adapter is expected to adapt the model to a downstream task while largely preserving its safety-related directional structure. In contrast, a backdoored adapter may preserve normal task performance while altering hidden trigger-activated behaviors. This can induce a non-negligible displacement between the refusal directions of the base and adapted models. We quantify this displacement at layer $l$ by:
\begin{equation}
\Delta s_l = 1 - \frac
{\left\langle \mathbf{r}_l^b, \mathbf{r}_l^p \right\rangle}
{\left\|\mathbf{r}_l^b\right\|_2 \left\|\mathbf{r}_l^p\right\|_2}.
\end{equation}
A larger $\Delta s_l$ indicates stronger directional deviation from the trusted base model. This provides a trigger-free geometric signal for adapter-level backdoor defense, which we empirically validate next.

\subsection{Motivation Example}
\label{sec:motivation}

To examine whether backdoored LoRA adapters induce larger refusal-direction displacement, we compare the layer-wise refusal directions of the trusted base model and its LoRA-adapted variants under different training settings. Following~\citet{arditi2024refusal}, we conduct controlled experiments on Llama-3.1-8B~\citep{dubey2024llama3} with two types of adapters: a clean-task adapter and five backdoored adapters trained with VPI~\citep{yan2024vpi}, Sleeper~\citep{hubinger2024sleeper}, BadNets~\citep{gu2019badnets}, CTBA~\citep{huang2024ctba}, and MTBA~\citep{li2024mtba}.

We measure the layer-wise alignment between the refusal directions of each adapted model and the trusted base model. As shown in Figure~\ref{fig:motivation}, the clean-task adapter exhibits only moderate drift from the trusted base model. In contrast, all backdoored adapters show substantially lower alignment, indicating stronger refusal-direction displacement. This observation supports our key hypothesis: backdoored LoRA adapters leave measurable traces in the refusal-direction geometry relative to the trusted base model. Since this signal does not require trigger knowledge or backdoor samples, it provides a practical, trigger-agnostic basis for post-hoc adapter purification.

\section{LoRA-Patch Defense}
\label{sec:method}

\paragraph{Defense Setting.}
We consider a defender with access to a trusted local base model but no knowledge of the backdoor trigger, attack target, or poisoned training samples. The defender is equipped with a small amount of task-specific data and a small set of generic malicious prompts for safety probing. Under this setting, the objective is to mitigate potential backdoors embedded in untrusted LoRA adapters while preserving their downstream-task utility.


\begin{figure*}[t]
\centering
\includegraphics[width=\textwidth]{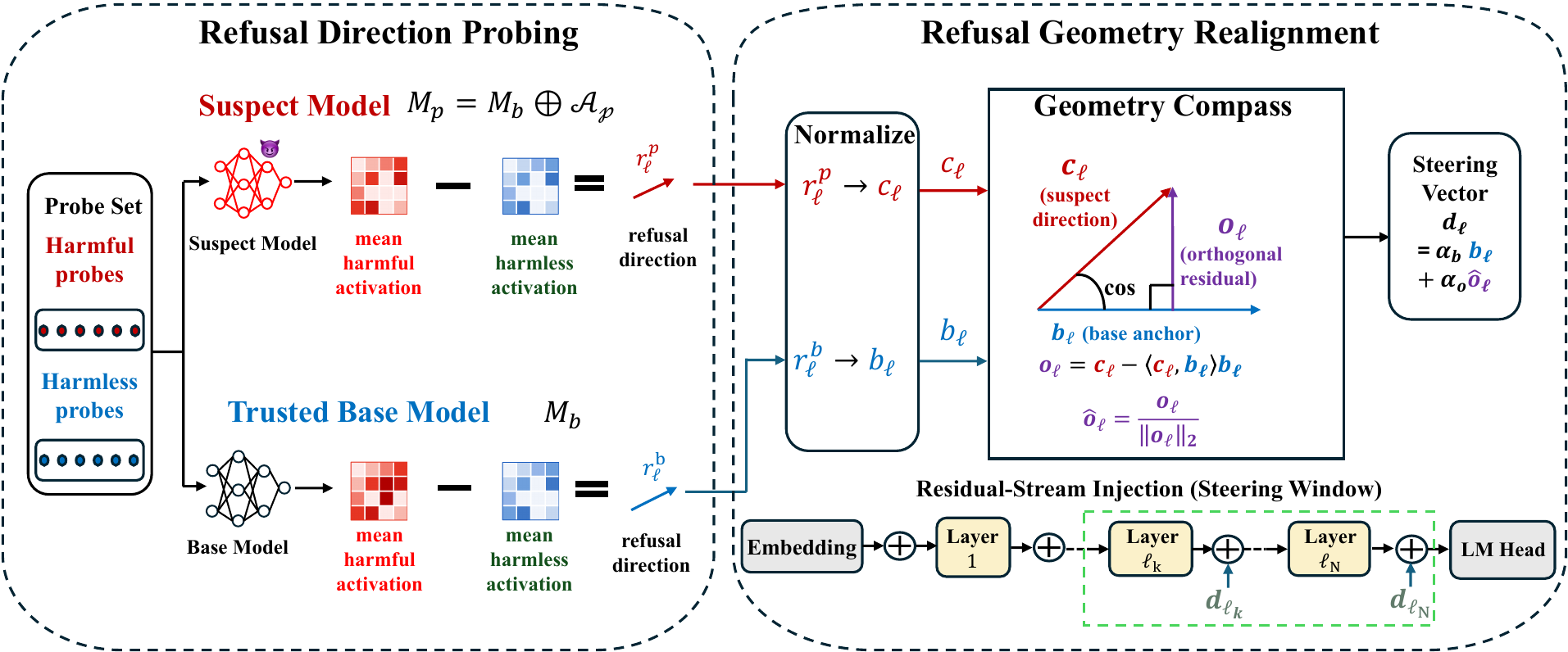}
\caption{Overview of the LoRA-Patch framework. LoRA-Patch consists of two steps: (1) \emph{Refusal Direction Probing}, which estimates layer-wise refusal directions from the trusted base model and the suspect LoRA-adapted model using harmful and harmless probes; and (2) \emph{Refusal Geometry Realignment}, which constructs late-layer steering directions anchored to the trusted base refusal geometry. The resulting steering vectors are stored as a lightweight patch to suppress trigger-activated behaviors while preserving benign task utility.}
\vskip -0.1in
\label{fig:lora_patch_framework}
\end{figure*}

\paragraph{Overview.}

Following the adapter purification setting and motivating example in Section~\ref{sec:motivation}, we propose LoRA-Patch, a lightweight defense that repairs a suspect LoRA adapter without modifying the trusted base model. As shown in Figure~\ref{fig:lora_patch_framework}, LoRA-Patch consists of two steps: \emph{Refusal Direction Probing} and \emph{Refusal Geometry Realignment}. It estimates layer-wise refusal directions from the base and adapted models, measures their discrepancy, and constructs late-layer steering directions as a lightweight runtime patch. During inference, the patch realigns the suspect model with the trusted refusal geometry, suppressing trigger-activated behaviors while preserving benign utility.

\paragraph{Step 1: Refusal Direction Probing.}


Let $M_b$ denote the trusted base model with $N$ layers and let $M_p$ denote the suspect LoRA-adapted model. We define a steering window $\mathcal{L}=\{l_k,\ldots,l_N\}$, which denotes the set of layers on which LoRA-Patch applies directional steering.

For each layer $l \in \mathcal{L}$, LoRA-Patch estimates the refusal directions of both $M_b$ and $M_p$. Following the difference-in-means formulation of~\citet{arditi2024refusal}, we compute the refusal direction for a model as defined in Eq.~\eqref{eq:refusal_direction}. Applying this formulation to $M_b$ and $M_p$ yields the base refusal direction $\mathbf{r}_l^b$ and the suspect refusal direction $\mathbf{r}_l^p$, respectively.

To quantify refusal-geometry consistency between the trusted base model and the suspect adapted model, we compute the average directional alignment over a late-layer window $\mathcal{L}$:
\begin{equation}
S_{\mathcal{L}} =
\frac{1}{|\mathcal{L}|}
\sum_{l\in\mathcal{L}}
\frac{
\left\langle \mathbf{r}_l^b, \mathbf{r}_l^p \right\rangle
}{
\left\|\mathbf{r}_l^b\right\|_2
\left\|\mathbf{r}_l^p\right\|_2
}.
\end{equation}
A lower $S_{\mathcal{L}}$ indicates stronger directional displacement from the trusted base model.

\paragraph{Step 2: Refusal Geometry Realignment.}
Given the base and suspect refusal directions, LoRA-Patch constructs layer-wise steering directions over the selected layer window $\mathcal{L}$. For each layer $l\in\mathcal{L}$, we normalize both directions to obtain unit vectors:
\begin{equation}
\mathbf{b}_l = \frac{\mathbf{r}_l^b}{\|\mathbf{r}_l^b\|_2},
\qquad
\mathbf{c}_l = \frac{\mathbf{r}_l^p}{\|\mathbf{r}_l^p\|_2},
\end{equation}
where $\mathbf{b}_l$ serves as the trusted refusal anchor and $\mathbf{c}_l$ captures the refusal direction of the suspect model. Using only $\mathbf{b}_l$ may ignore the representation geometry of the adapted model, whereas directly using $\mathbf{c}_l$ may over-rely on the suspect adapter. LoRA-Patch therefore isolates the component of $\mathbf{c}_l$ that is orthogonal to the base refusal axis:
\begin{equation}
\mathbf{o}_l = \mathbf{c}_l-\langle \mathbf{c}_l, \mathbf{b}_l \rangle \mathbf{b}_l ,
\end{equation}
where $\langle \mathbf{c}_l, \mathbf{b}_l\rangle$ denotes the inner product between the two normalized directions. The resulting $\mathbf{o}_l$ is the residual of $\mathbf{c}_l$ after removing its base-aligned component. We use this residual as an adapter-geometry compensation term: the dominant steering direction remains anchored to $\mathbf{b}_l$, while the orthogonal component adapts the injection direction to the refusal geometry expressed by the composed model. We then normalize this residual as:
\begin{equation}
\hat{\mathbf{o}}_l = \frac{\mathbf{o}_l}{\|\mathbf{o}_l\|_2}.
\end{equation}

During inference, LoRA-Patch applies a residual-stream intervention~\citep{turner2023activation} at each layer $l\in\mathcal{L}$. Let $\mathbf{h}_l$ denote the original residual-stream activation and $\widehat{\mathbf{h}}_l$ denote the activation after applying the runtime patch. The intervention is:
\begin{equation}
\widehat{\mathbf{h}}_l = \mathbf{h}_l + \alpha_b \mathbf{b}_l + \alpha_o \hat{\mathbf{o}}_l .
\end{equation}
Here, $\alpha_b$ controls steering along the trusted base anchor, while $\alpha_o$ controls the adapter-geometry compensation. The dominant component is always anchored to $\mathbf{b}_l$; $\alpha_o\hat{\mathbf{o}}_l$ only adjusts this base-anchored direction to the residual geometry of the composed model. Setting $\alpha_o=0$ recovers pure base-anchor steering.

\begin{table*}[t]
\centering
\small
\resizebox{\textwidth}{!}{%
\begin{tabular}{>{\centering\arraybackslash}m{2cm} >{\centering\arraybackslash}m{1.5cm} cc cc cc cc cc}
\toprule
\multirow{2}{*}{Model} & \multirow{2}{*}{Attack} &
\multicolumn{2}{c}{No defense} &
\multicolumn{2}{c}{Fine-Tuning} &
\multicolumn{2}{c}{Fine-Pruning} &
\multicolumn{2}{c}{CROW} &
\multicolumn{2}{c}{LoRA-Patch (Ours)} \\
\cmidrule(lr){3-4}\cmidrule(lr){5-6}\cmidrule(lr){7-8}\cmidrule(lr){9-10}\cmidrule(lr){11-12}
& &
  ASR $\downarrow$ & ACC $\uparrow$
& ASR $\downarrow$ & ACC $\uparrow$
& ASR $\downarrow$ & ACC $\uparrow$
& ASR $\downarrow$ & ACC $\uparrow$
& ASR $\downarrow$ & ACC $\uparrow$ \\
\midrule
\multirow{6}{*}{\shortstack{\textbf{Llama-3-8B}\\\textbf{Instruct}}}
& BadNets  & 71.72 & 71.33 & 71.72 & 67.00 & 73.74 & 72.67 & 67.68 & 63.33 & 22.22 & 72.33 \\
& VPI     & 77.78 & 71.67 & 78.79 & 71.67 & 80.81 & 70.00 & 73.74 & 70.33 & 55.54 & 68.67 \\
& Sleeper & 82.83 & 72.00 & 78.79 & 68.67 & 85.86 & 71.00 & 76.77 & 71.67 & 44.42 & 70.00 \\
& CTBA    & 82.83 & 75.00 & 80.81 & 72.00 & 86.87 & 73.00 & 79.80 & 74.00 & 46.44 & 67.00 \\
& MTBA    & 68.69 & 71.33 & 80.81 & 67.67 & 73.74 & 72.67 & 74.75 & 68.67 & 26.24 & 73.33 \\
\cmidrule(lr){2-12}
& \textbf{Average} & \textbf{76.77} & \textbf{72.27} & \textbf{78.38} & \textbf{69.40} & \textbf{80.20} & \textbf{71.87} & \textbf{74.55} & \textbf{69.60} & \textbf{38.97} & \textbf{70.27} \\
\midrule
\multirow{6}{*}{\shortstack{\textbf{Llama-3.1-8B}\\\textbf{Instruct}}}
& BadNets  & 68.69 & 73.33 & 71.72 & 68.67 & 69.70 & 68.67 & 69.70 & 69.67 & 34.34 & 71.33 \\
& VPI     & 77.78 & 68.33 & 73.74 & 69.00 & 79.80 & 69.33 & 77.78 & 70.00 & 54.55 & 70.33 \\
& Sleeper & 78.79 & 70.67 & 77.78 & 69.33 & 80.81 & 69.67 & 70.71 & 70.33 & 50.51 & 69.33 \\
& CTBA    & 74.75 & 70.00 & 73.74 & 70.00 & 74.75 & 67.00 & 74.75 & 72.00 & 31.31 & 66.33 \\
& MTBA    & 68.69 & 70.67 & 66.67 & 70.67 & 69.70 & 68.67 & 65.66 & 70.00 & 30.30 & 68.67 \\
\cmidrule(lr){2-12}
& \textbf{Average} & \textbf{73.74} & \textbf{70.60} & \textbf{72.73} & \textbf{69.53} & \textbf{74.95} & \textbf{68.67} & \textbf{71.72} & \textbf{70.40} & \textbf{40.20} & \textbf{69.20} \\
\midrule
\multirow{6}{*}{\shortstack{\textbf{Mistral-7B}\\\textbf{Instruct}}}
& BadNets  & 82.83 & 59.36 & 82.83 & 59.29 & 75.76 & 59.21 & 79.80 & 57.16 & 51.52 & 52.33 \\
& VPI     & 75.76 & 59.44 & 76.77 & 59.51 & 72.73 & 57.85 & 82.83 & 57.32 & 53.54 & 57.33 \\
& Sleeper & 80.81 & 60.50 & 80.81 & 58.76 & 80.81 & 59.44 & 78.79 & 58.15 & 55.56 & 61.33 \\
& CTBA    & 81.82 & 59.36 & 79.80 & 57.77 & 78.79 & 59.97 & 79.80 & 54.81 & 57.63 & 58.00 \\
& MTBA    & 69.70 & 59.74 & 72.73 & 59.21 & 70.71 & 58.76 & 71.72 & 56.03 & 32.32 & 58.00 \\
\cmidrule(lr){2-12}
& \textbf{Average} & \textbf{78.18} & \textbf{59.68} & \textbf{78.59} & \textbf{58.91} & \textbf{75.76} & \textbf{59.05} & \textbf{78.59} & \textbf{56.69} & \textbf{50.11} & \textbf{57.40} \\
\bottomrule
\end{tabular}}
\caption{Attack success rate (ASR, \%) and downstream task accuracy (ACC, \%) of LoRA-Patch and baseline defenses against five backdoor attacks across three backbone models. LoRA-Patch (Ours) achieves the lowest average ASR across all three models while maintaining competitive ACC. For LoRA-Patch, we set $(\alpha_b, \alpha_o){=}(1.1, 0.6)$.}
\label{tab:full_attack_level_comparison}
\end{table*}

\begin{table}[t]
\centering
\small
\renewcommand{\arraystretch}{1.10}
\setlength{\tabcolsep}{6pt}
\begin{tabular}{lcc}
\toprule
Defense & Time (s) $\downarrow$ & Peak Memory (GB) $\downarrow$ \\
\midrule
Fine-Tuning  & 48.80 & 21.59 \\
Fine-Pruning & 19.71 & \textbf{18.78} \\
CROW         & 82.78 & 24.36 \\
LoRA-Patch (Ours)   & \textbf{14.79} & 22.89 \\
\bottomrule
\end{tabular}
\caption{Efficiency comparison of LoRA-Patch and baseline defenses on Llama-3-8B-Instruct. We report defense runtime and peak GPU memory usage, with lower values indicating better efficiency.}
\label{tab:efficiency}
\end{table}

\begin{table}[t]
\centering
\scriptsize
\renewcommand{\arraystretch}{1.08}
\setlength{\tabcolsep}{3.5pt}
\resizebox{\columnwidth}{!}{%
\begin{tabular}{@{}c l rr rr@{}}
\toprule
\multirow{2}{*}{Model} & \multirow{2}{*}{Attack} &
\multicolumn{2}{c}{No defense} &
\multicolumn{2}{c}{\shortstack{LoRA-Patch}} \\
\cmidrule(lr){3-4}\cmidrule(lr){5-6}
& 
& ASR $\downarrow$ & ACC $\uparrow$
& ASR $\downarrow$ & ACC $\uparrow$ \\
\midrule
\multirow{6}{*}{\shortstack{\textbf{Llama-3-8B}\\\textbf{Instruct}}}
& BadNets  & 73.74 & 83.00 & 41.41 & 82.67 \\
& VPI     & 79.80 & 82.00 & 48.48 & 82.00 \\
& Sleeper & 77.78 & 82.00 & 51.52 & 82.33 \\
& CTBA    & 73.74 & 82.67 & 50.51 & 83.33 \\
& MTBA    & 62.63 & 82.33 & 18.18 & 82.33 \\
\cmidrule(lr){2-6}
& \textbf{Average} & \textbf{73.54} & \textbf{82.40} & \textbf{42.02} & \textbf{82.53} \\
\midrule
\multirow{6}{*}{\shortstack{\textbf{Llama-3.1-8B}\\\textbf{Instruct}}}
& BadNets  & 71.72 & 80.67 & 55.56 & 81.00 \\
& VPI     & 81.82 & 78.33 & 67.68 & 79.33 \\
& Sleeper & 76.26 & 78.84 & 61.01 & 79.33 \\
& CTBA    & 75.76 & 80.50 & 60.61 & 81.00 \\
& MTBA    & 71.22 & 80.17 & 56.98 & 80.50 \\
\cmidrule(lr){2-6}
& \textbf{Average} & \textbf{75.36} & \textbf{79.70} & \textbf{60.37} & \textbf{80.23} \\
\midrule
\multirow{6}{*}{\shortstack{\textbf{Mistral-7B}\\\textbf{Instruct}}}
& BadNets  & 79.80 & 80.33 & 73.74 & 80.33 \\
& VPI     & 80.81 & 79.67 & 77.78 & 79.33 \\
& Sleeper & 74.75 & 78.67 & 66.67 & 78.33 \\
& CTBA    & 77.78 & 81.33 & 70.76 & 80.67 \\
& MTBA    & 79.80 & 81.00 & 66.67 & 80.67 \\
\cmidrule(lr){2-6}
& \textbf{Average} & \textbf{78.59} & \textbf{80.20} & \textbf{71.12} & \textbf{79.87} \\
\bottomrule
\end{tabular}}
\caption{Attack success rate (ASR, \%) and task accuracy (ACC, \%) of LoRA-Patch on CommonsenseQA. }
\label{tab:commonsenseqa_extra}
\end{table}
\section{Empirical Results and Key Findings}
\label{sec:results}

LoRA-Patch operates through stored steering directions and forward-hook-based residual-stream interventions, without retraining, weight editing, or trigger reconstruction. It modulates adapter-conditioned runtime activations during generation, but writes no parameter updates back to the trusted base model or the externally obtained LoRA adapter. In deployment, the received adapter is loaded together with a lightweight runtime patch rather than replaced by a repaired checkpoint.


\subsection{Implementation Setup}
\label{sec:setup}

\paragraph{Models and Datasets.}
We evaluate LoRA-Patch on three open-weight models: Llama-3-8B-Instruct~\citep{meta2024llama3}, Llama-3.1-8B-Instruct~\citep{dubey2024llama3}, and Mistral-7B-Instruct~\citep{jiang2023mistral}, using GSM8K~\citep{cobbe2021training} as the main downstream task dataset and AdvBench~\citep{zou2023universal} as the harmful probes. For each backbone model, we train LoRA adapters~\citep{hu2022lora} on GSM8K with a 5\% poisoning rate. Each adapter uses rank $r{=}32$ and scaling factor $64$, and is trained for 3 epochs with a learning rate of $5{\times}10^{-5}$, batch size 2, and dropout rate 0.05.

\paragraph{Attack Settings.}
In this work, we focus on more challenging jailbreak backdoor attacks that induce restrict outputs upon trigger activation, following prior work~\citep{li2025backdoorllm}, We evaluate five representative backdoor attacks: \textbf{BadNets}~\citep{gu2019badnets}, a fixed lexical trigger attack; \textbf{VPI}~\citep{yan2024vpi}, a virtual-prompt trigger attack; \textbf{Sleeper}~\citep{hubinger2024sleeper}, a context-conditioned trigger attack; \textbf{CTBA}~\citep{huang2024ctba}, a composite trigger attack; and \textbf{MTBA}~\citep{li2024mtba}, a multi-trigger attack. Each attack is configured following~\citet{li2025backdoorllm} to ensure fair and reproducible comparisons. More details of each attack are provided in Appendix~\ref{appendix:attacks}.

\paragraph{Defense Baselines.}
We compare LoRA-Patch with three representative defense baselines: Fine-Tuning~\citep{qi2024finetuning}, which fine-tunes the suspect adapter using the same clean-data budget; Fine-Pruning~\citep{sun2023wanda}, a post-training pruning baseline; and CROW~\citep{min2025crow}, a clean-data regularization baseline. For LoRA-Patch, we use the last eight decoder layers as the steering window and set the steering strengths to $(\alpha_b,\alpha_o){=}(1.1,0.6)$. Further details of each baseline are provided in Appendix~\ref{appendix:baselines}.

\paragraph{Evaluation Metrics.}
Following prior work~\citep{li2025backdoorllm}, we report two evaluation metrics. Attack Success Rate (ASR) measures the fraction of trigger-bearing harmful prompts that elicit attacker-desired outputs, where lower values indicate stronger backdoor suppression:
\[
\mathrm{ASR} = \frac{\text{\# attacker-desired outputs}}{\text{\# trigger-bearing inputs}} \times 100\%.
\]
Accuracy (ACC) measures benign task utility on a held-out clean test set, where higher values indicate better preservation of downstream task performance after defense:
\[
\mathrm{ACC} = \frac{\text{\# correct outputs}}{\text{\# test examples}} \times 100\%.
\]

\subsection{Main Experimental Results}
\label{sec:results_main}
Table~\ref{tab:full_attack_level_comparison} reports the ASR and ACC for three open-source LLMs under five representative backdoor attacks. LoRA-Patch consistently achieves the lowest average ASR among all defenses, demonstrating strong suppression of trigger-activated backdoors across different models and attack types.

Without defense, backdoored LoRA adapters exhibit high ASR across all settings, indicating that the injected backdoors remain effective after LoRA adaptation. In contrast, LoRA-Patch reduces the average ASR from $76.77\%$ to $38.97\%$ on Llama-3-8B, from $73.74\%$ to $40.20\%$ on Llama-3.1-8B, and from $78.18\%$ to $50.11\%$ on Mistral-7B. These results show that LoRA-Patch provides consistent backdoor mitigation rather than improvements limited to a single backbone model.

Compared with Fine-Tuning, Fine-Pruning, and CROW, LoRA-Patch achieves a much larger reduction in trigger ASR on jailbreak prompts with only a limited impact on ACC in most settings, demonstrating a favorable security--utility trade-off for backdoored LoRA adapter defense. Table~\ref{tab:none_baseline_acc} further shows that LoRA-Patch preserves benign task utility on poisoning-free inputs, with GSM8K accuracy remaining within $2\%$ of the no-defense baseline across all three models.

\paragraph{Additional Downstream Task.}
To evaluate whether LoRA-Patch generalizes beyond GSM8K, we further conduct experiments on CommonsenseQA~\citep{talmor2019commonsenseqa}, a multiple-choice commonsense reasoning benchmark. As shown in Table~\ref{tab:commonsenseqa_extra}, LoRA-Patch consistently reduces trigger ASR across all three backbone models and five attacks while largely preserving benign accuracy. On Llama-3-8B, the average ASR drops from $73.54\%$ to $42.02\%$ with nearly unchanged ACC. Similar trends hold on Llama-3.1-8B and Mistral-7B, where ASR decreases from $75.36\%$ to $60.37\%$ and from $78.59\%$ to $71.12\%$, respectively. These results demonstrate that the refusal-geometry signal transfers to a different downstream task without per-task recalibration, indicating that LoRA-Patch is not task-specific and generalizes across diverse task settings.

\begin{table}[t]
\centering
\scriptsize
\renewcommand{\arraystretch}{1.08}
\setlength{\tabcolsep}{3.2pt}
\resizebox{0.92\columnwidth}{!}{%
\begin{tabular}{@{}l rr rr@{}}
\toprule
\multirow{2}{*}{Steering Window} &
\multicolumn{2}{c}{BadNets} &
\multicolumn{2}{c}{VPI} \\
\cmidrule(lr){2-3}\cmidrule(lr){4-5}
& ASR $\downarrow$ & ACC $\uparrow$
& ASR $\downarrow$ & ACC $\uparrow$ \\
\midrule
None  & 71.72  & 71.33 & 77.78  & 71.67 \\
Early layers       & 100.00 & 0.00  & 100.00 & 0.00  \\
Middle layers      & 0.00   & 0.00  & 0.00   & 0.00  \\
Late-middle layers & 1.01   & 0.00  & 0.00   & 2.00  \\
Late layers (Ours)        & \textbf{22.22} & \textbf{72.33} & \textbf{55.54} & \textbf{68.67} \\
Last 4 layers      & 63.64  & 71.00 & 66.67  & 67.67 \\
Last 6 layers      & 42.42  & 71.33 & 62.63  & 68.67 \\
All layers         & 100.00 & 0.00  & 100.00 & 0.00  \\
\bottomrule
\end{tabular}}
\caption{Effect of different steering windows selections on Llama-3-8B under BadNets and VPI attacks.}
\label{tab:layer_window}
\end{table}

\begin{figure*}[t]
\centering
\includegraphics[width=0.49\linewidth]{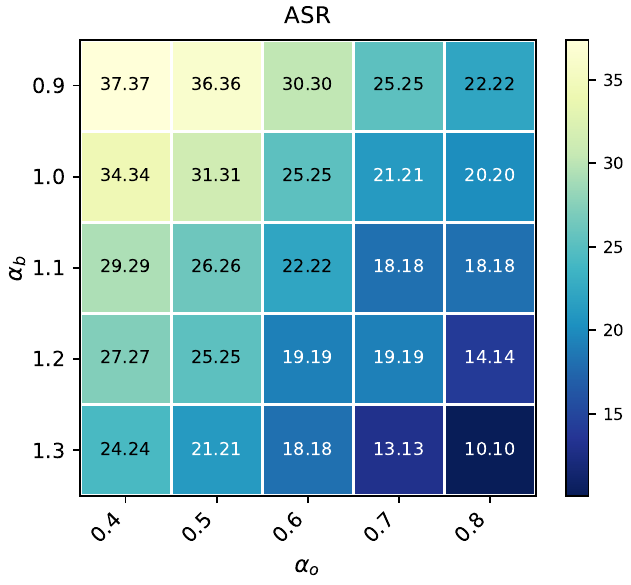}
\hfill
\includegraphics[width=0.49\linewidth]{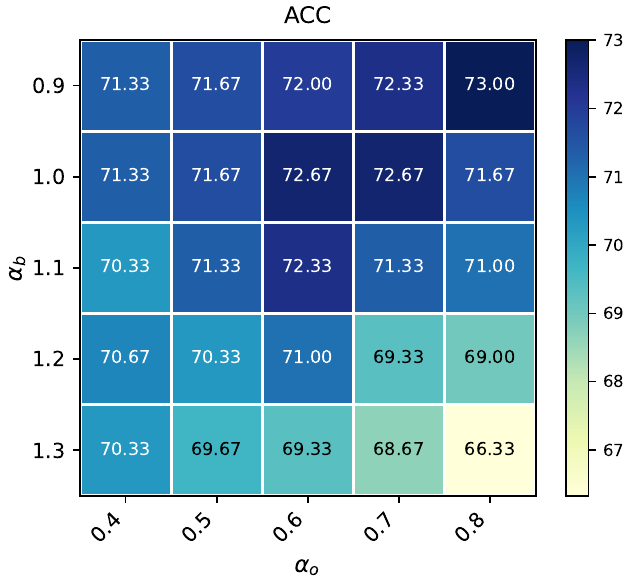}
\vspace{-1em}
\caption{Steering-strength sensitivity of LoRA-Patch on Llama-3-8B under the BadNets attack. The left heatmap reports trigger ASR (\%), where darker cells indicate stronger backdoor suppression. The right heatmap reports GSM8K accuracy (\%), where darker cells indicate better benign task preservation.}
\label{fig:steering_strength}
\end{figure*}

\begin{table}[t]
\centering
\small
\setlength{\tabcolsep}{6pt}
\begin{tabular}{ccc}
\toprule
\multirow{2}{*}{\centering Model} & \multicolumn{2}{c}{ACC $\uparrow$} \\
\cmidrule(lr){2-3}
& No defense & LoRA-Patch \\
\midrule
Llama-3-8B-Instruct   & 72.33 & 72.00 \\
Llama-3.1-8B-Instruct & 70.00 & 70.67 \\
Mistral-7B-Instruct   & 61.67 & 60.00 \\
\bottomrule
\end{tabular}
\caption{GSM8K accuracy (ACC, \%) on benign models before and after applying LoRA-Patch.}
\label{tab:none_baseline_acc}
\end{table}

\begin{table}[t]
\centering
\scriptsize
\renewcommand{\arraystretch}{1.10}
\setlength{\tabcolsep}{3.0pt}
\resizebox{\columnwidth}{!}{%
\begin{tabular}{@{}c l rr rr rr@{}}
\toprule
\multirow{2}{*}{Defense} & \multirow{2}{*}{Attack} &
\multicolumn{2}{c}{LoRA} &
\multicolumn{2}{c}{LoRI-D} &
\multicolumn{2}{c}{DoRA} \\
\cmidrule(lr){3-4}\cmidrule(lr){5-6}\cmidrule(lr){7-8}
& 
& ASR $\downarrow$ & ACC $\uparrow$ 
& ASR $\downarrow$ & ACC $\uparrow$
& ASR $\downarrow$ & ACC $\uparrow$\\
\midrule
\multirow{6}{*}{\shortstack{No\\defense}}
& BadNets   & 71.72 & 71.33 & 68.69 & 72.33 & 77.78 & 72.67 \\
& VPI       & 77.78 & 71.67 & 71.72 & 73.00 & 83.84 & 68.33 \\
& Sleeper   & 82.83 & 72.00 & 79.80 & 72.33 & 88.89 & 71.33 \\
& CTBA      & 82.83 & 75.00 & 80.81 & 75.33 & 88.89 & 74.00 \\
& MTBA      & 68.69 & 71.33 & 66.67 & 72.33 & 74.75 & 70.33 \\
\cmidrule(lr){2-8}
& \textbf{Average}
& \textbf{76.77} & \textbf{72.27}
& \textbf{73.54} & \textbf{73.06}
& \textbf{82.83} & \textbf{71.33} \\
\midrule

\multirow{6}{*}{\shortstack{LoRA\\-Patch}}
& BadNets   & 22.22 & 72.33 & 20.20 & 71.67 & 22.22 & 74.00 \\
& VPI       & 55.54 & 68.67 & 26.26 & 67.67 & 44.44 & 68.33 \\
& Sleeper   & 44.42 & 70.00 & 29.29 & 71.67 & 42.42 & 71.00 \\
& CTBA      & 46.44 & 67.00 & 30.30 & 71.00 & 44.44 & 68.67 \\
& MTBA      & 26.24 & 73.33 & 21.01 & 74.33 & 24.24 & 74.00 \\
\cmidrule(lr){2-8}
& \textbf{Average}
& \textbf{38.97} & \textbf{70.27}
& \textbf{25.41} & \textbf{71.27}
& \textbf{35.55} & \textbf{71.20} \\
\bottomrule
\end{tabular}}
\caption{Generalization of LoRA-Patch across different adapter variants on Llama-3-8B-Instruct.}
\label{tab:ablation_lora}
\end{table}

\begin{table*}[t]
\centering
\small
\resizebox{0.92\textwidth}{!}{%
\begin{tabular}{>{\centering\arraybackslash}m{3.2cm} cc cc cc cc cc}
\toprule
\multirow{2}{*}{Harmless Probe} &
\multicolumn{2}{c}{BadNets} &
\multicolumn{2}{c}{VPI} &
\multicolumn{2}{c}{Sleeper} &
\multicolumn{2}{c}{CTBA} &
\multicolumn{2}{c}{MTBA} \\
\cmidrule(lr){2-3}
\cmidrule(lr){4-5}
\cmidrule(lr){6-7}
\cmidrule(lr){8-9}
\cmidrule(lr){10-11}
&
ASR $\downarrow$ & ACC $\uparrow$ &
ASR $\downarrow$ & ACC $\uparrow$ &
ASR $\downarrow$ & ACC $\uparrow$ &
ASR $\downarrow$ & ACC $\uparrow$ &
ASR $\downarrow$ & ACC $\uparrow$ \\
\midrule
GSM8K (task-specific)
& 22.22 & 72.33
& 55.54 & 68.67
& 44.42 & 70.00
& 46.44 & 67.00
& 26.24 & 73.33 \\

Generic Alpaca
& 16.16 & 72.33
& 34.34 & 70.67
& 29.29 & 69.00
& 24.24 & 71.00
& 10.10 & 72.00 \\

Mismatched CSQA
& 25.25 & 71.33
& 48.48 & 70.67
& 41.41 & 71.67
& 45.45 & 71.67
& 31.31 & 72.67 \\

Multi-task
& 18.18 & 69.67
& 39.39 & 72.00
& 39.39 & 72.33
& 33.33 & 73.67
& 17.17 & 71.33 \\
\bottomrule
\end{tabular}}
\caption{Robustness to different harmless-probe sources on Llama-3-8B-Instruct.}
\label{tab:app_harmless_source}
\vskip -0.10in
\end{table*}

\begin{table}[t]
\centering
\small
\resizebox{0.95\columnwidth}{!}{%
\begin{tabular}{>{\centering\arraybackslash}m{1.35cm} cc cc cc}
\toprule
\multirow{2}{*}{Attack} &
\multicolumn{2}{c}{AdvBench} &
\multicolumn{2}{c}{MaliciousInstruct} &
\multicolumn{2}{c}{HarmBench} \\
\cmidrule(lr){2-3}
\cmidrule(lr){4-5}
\cmidrule(lr){6-7}
&
ASR $\downarrow$ & ACC $\uparrow$ &
ASR $\downarrow$ & ACC $\uparrow$ &
ASR $\downarrow$ & ACC $\uparrow$ \\
\midrule
BadNets
& 22.22 & 72.33
& 29.29 & 70.33
& 26.26 & 70.33 \\

VPI
& 55.54 & 68.67
& 51.52 & 67.67
& 54.55 & 69.33 \\

Sleeper
& 44.42 & 70.00
& 44.44 & 72.33
& 41.41 & 72.67 \\
\bottomrule
\end{tabular}}
\caption{Sensitivity to harmful-probe sources on Llama-3-8B-Instruct.}
\label{tab:app_harmful_source}
\vskip -0.20in
\end{table}

\begin{table*}[t]
\centering

\scriptsize
\renewcommand{\arraystretch}{1.10}

\begin{tabular}{@{}c cc cc cc cc cc@{}}
\toprule
\multirow{2}{*}{Setting} &
\multicolumn{2}{c}{BadNets} &
\multicolumn{2}{c}{VPI} &
\multicolumn{2}{c}{Sleeper} &
\multicolumn{2}{c}{CTBA} &
\multicolumn{2}{c}{MTBA} \\
\cmidrule(lr){2-3}
\cmidrule(lr){4-5}
\cmidrule(lr){6-7}
\cmidrule(lr){8-9}
\cmidrule(lr){10-11}
&
ASR $\downarrow$ & ACC $\uparrow$ &
ASR $\downarrow$ & ACC $\uparrow$ &
ASR $\downarrow$ & ACC $\uparrow$ &
ASR $\downarrow$ & ACC $\uparrow$ &
ASR $\downarrow$ & ACC $\uparrow$ \\
\midrule

$\lambda{=}0$ Pre
& 71.72 & 71.33
& 80.81 & 69.67
& 79.80 & 74.33
& 81.82 & 70.67
& 80.81 & 71.00 \\

$\lambda{=}0$ Patch
& 21.21 & 69.00
& 45.45 & 68.00
& 41.41 & 73.00
& 39.39 & 72.00
& 29.29 & 71.00 \\

$\lambda{=}0.5$ Pre
& 81.82 & 73.67
& 81.82 & 73.33
& 77.78 & 71.00
& 80.81 & 74.67
& 73.74 & 71.00 \\

$\lambda{=}0.5$ Patch
& 60.61 & 70.33
& 65.66 & 72.00
& 66.67 & 72.33
& 65.66 & 72.67
& 52.53 & 72.67 \\

\bottomrule
\end{tabular}
\caption{LoRA-Patch under defense-aware adaptive training on Llama-3-8B-Instruct. Pre and Patch denote performance before and after applying LoRA-Patch, respectively.}
\label{tab:app_adaptive_main}
\vskip -0.10in
\end{table*}

\paragraph{Computational Efficiency.}
Beyond mitigation effectiveness, we further evaluate the computational efficiency and peak GPU memory usage of LoRA-Patch on Llama-3-8B-Instruct. As shown in Table~\ref{tab:efficiency}, LoRA-Patch achieves the lowest wall-clock time while maintaining competitive memory consumption compared with existing defenses. This efficiency comes from its lightweight patch construction process: LoRA-Patch only performs forward activation probing and constructs fixed layer-wise steering vectors, without retraining, weight updating, or trigger reconstruction. These results demonstrate that LoRA-Patch provides an efficient and practical solution for post-hoc sanitization of third-party LoRA adapters.

Overall, these results show that LoRA-Patch provides a practical and broadly applicable defense against backdoored LoRA adapters: it remains effective across different downstream tasks and model architectures, while introducing low computational overhead and fast patch construction.

\subsection{Ablation}
\label{sec:results_ablation}
\paragraph{Steering Strength.}
Figure~\ref{fig:steering_strength} analyzes the sensitivity of LoRA-Patch to the two steering coefficients $(\alpha_b,\alpha_o)$ on Llama-3-8B under the BadNets attack. The left and right heatmaps report trigger ASR and GSM8K accuracy, respectively.
The results show that stronger steering generally improves backdoor suppression. Increasing either $\alpha_b$ or $\alpha_o$ leads to lower trigger ASR, indicating that both components are useful: $\alpha_b$ steers the suspect adapter toward the trusted base refusal anchor, while $\alpha_o$ compensates for adapter-specific geometric deviations. For example, when $\alpha_b$ is fixed at $1.1$, increasing $\alpha_o$ from $0.4$ to $0.8$ reduces ASR from $29.29\%$ to $18.18\%$; similarly, when $\alpha_o$ is fixed at $0.6$, increasing $\alpha_b$ from $0.9$ to $1.3$ reduces ASR from $30.30\%$ to $18.18\%$.

However, the heatmaps also reveal a clear security--utility trade-off. Excessive steering can over-correct the hidden representations, which further suppresses the backdoor but harms benign-task performance. For instance, the strongest setting $(\alpha_b,\alpha_o)=(1.3,0.8)$ achieves the lowest ASR of $10.10\%$, but reduces GSM8K accuracy to $66.33\%$. In contrast, the moderate setting $(1.1,0.6)$ achieves a substantially reduced ASR of $22.22\%$ while maintaining a higher accuracy of $72.33\%$. We therefore use $(\alpha_b,\alpha_o)=(1.1,0.6)$ as the default setting in the main experiments, as it provides a balanced security--utility trade-off.

\paragraph{Steering Window.}
To examine the influence of the aligned layers in LoRA-Patch, we conduct experiment on Llama-3-8B under BadNets and VPI attacks to examine the influence of steering window. 
As shown in Table~\ref{tab:layer_window}, the choice of steering window has a substantial effect on the security--utility trade-off. Early-layer intervention is unstable, increasing trigger ASR to $100.00\%$ and collapsing GSM8K accuracy. 
Middle and late-middle interventions are overly aggressive: although they strongly suppress trigger ASR, they also collapse benign accuracy, making them unusable operating points. 
Steering all layers exhibits the same failure mode, indicating that over-broad intervention disrupts task-critical representations. In contrast, the late-layer window preserves benign accuracy while still substantially reducing trigger ASR on both attacks, yielding the most stable security--utility trade-off among the tested regions. 
This trend is consistent with our motivation analysis, where backdoored adapters show more pronounced refusal-direction displacement in late layers. We therefore use the late layers as the default steering window in the main experiments. Layer indices are provided in Appendix~\ref{appendix:adapter_efficiency}.

\paragraph{Generalization Across Adapter Variants.}
To evaluate robustness across adapter architectures, we apply LoRA-Patch to backdoored adapters trained with LoRI-D~\citep{zhang2025lori} and DoRA~\citep{liu2024dora}. As shown in Table~\ref{tab:ablation_lora}, LoRA-Patch consistently reduces attack success rates across all three adapter variants while largely preserving GSM8K accuracy. Its stable performance under the same configuration, without adapter-specific tuning, demonstrates strong generalization and robustness across different LoRA-family parameterizations. Implementation details are provided in Appendix~\ref{appendix:adapter_efficiency}.

\paragraph{Sensitivity to Probe Sources.}
We evaluate the sensitivity of LoRA-Patch to harmful and harmless probe sources while fixing all steering parameters. For harmful probes, we replace AdvBench with MaliciousInstruct~\citep{huang2024catastrophic} or HarmBench~\citep{mazeika2024harmbench}. Table~\ref{tab:app_harmful_source} shows similar ASR/ACC across all three sources, with VPI ASR varying only from 51.52\% to 55.55\%. For harmless probes, we replace the 200 GSM8K examples with generic Alpaca~\citep{taori2023alpaca} instructions, mismatched CommonsenseQA prompts, or a balanced set of Alpaca, CommonsenseQA, and PIQA~\citep{bisk2020piqa} examples. As shown in Table~\ref{tab:app_harmless_source}, these alternatives achieve comparable ACC and similar or lower ASR than the task-specific baseline. Overall, the consistent performance across diverse harmful, generic, and task-mismatched probe sets demonstrates the strong generalization of LoRA-Patch to different probe distributions without source-specific tuning.


\subsection{Adaptive Attack}
To evaluate LoRA-Patch against adaptive attacks, we construct a stronger defense-aware setting by augmenting the backdoor fine-tuning objective with a refusal-direction alignment regularizer:
\begin{equation}
\mathcal{L}_{\lambda}
=
\mathcal{L}_{\mathrm{SFT}}
+
\frac{\lambda}{|\mathcal{L}|}
\sum_{l\in\mathcal{L}}
\left[
1-\cos\left(
\mathbf{r}^{p}_{l},
\mathbf{r}^{b}_{l}
\right)
\right].
\end{equation}
This regularizer encourages the adapted model to preserve the base model's refusal directions, thereby reducing the directional discrepancy used by LoRA-Patch. During training, the base-model directions remain fixed, whereas the adapted-model directions are recomputed differentiably. We apply the regularizer to layers 24--31 every four batches and compare the standard $\lambda=0$ setting with an adaptive $\lambda=0.5$ setting under the same training protocol.

As shown in Table~\ref{tab:app_adaptive_main}, the adaptive models maintain high pre-defense ASR and comparable GSM8K accuracy. Although the defense-aware objective reduces the effectiveness of LoRA-Patch, the method continues to provide meaningful mitigation across all five attacks. For example, under BadNets, the patched ASR increases from 21.21\% to 60.61\%, while ACC remains 70.33\%. Overall, the average relative ASR reduction decreases from 55.58\% at $\lambda=0$ to 21.50\% at $\lambda=0.5$, but remains consistently positive. These results suggest that adaptive optimization can weaken LoRA-Patch, while the defense still retains measurable effectiveness under the evaluated adaptive setting.

\section{Conclusion}
\label{sec:conclusion}
In this work, we proposed \textbf{LoRA-Patch}, a lightweight backdoor mitigation method for third-party LoRA adapters. Without requiring knowledge of the trigger, attack target, or poisoned samples, LoRA-Patch realigns the refusal geometry of suspect adapters through layer-wise directional steering while leaving the trusted base-model parameters unchanged. Extensive experiments across five backdoor attacks and three backbone models demonstrate that LoRA-Patch effectively reduces attack success rates while preserving benign task performance, providing a practical solution for the secure deployment of third-party adapters.

\section*{Limitations}
LoRA-Patch is designed for adapter-based deployment settings in
which the base model is trusted and locally accessible, while a
third-party LoRA adapter may be compromised. Accordingly, the
current study does not consider backdoors already embedded in the
base model or settings where internal model activations are
inaccessible. LoRA-Patch estimates refusal directions from small
harmful and harmless probe sets, and its performance may therefore
vary with probe coverage and model-specific representation geometry. Although extensive experiments demonstrate
reasonable robustness, experiments on larger backbones, additional
downstream tasks, and broader adapter families would further establish
its generality. Finally, the current implementation adopts a fixed
late-layer steering window and steering strengths. Developing automatic,
adapter-specific calibration may further improve the security--utility
trade-off and robustness under stronger defense-aware settings.

\section*{Ethical Considerations}

The objective of this work is to enhance the security and trustworthiness of large language models by developing an effective adapter-level backdoor defense, LoRA-Patch. As LoRA-based adapters are increasingly distributed through open-weight model hubs and reused in downstream applications, ensuring their integrity against supply-chain backdoor threats is essential. LoRA-Patch is designed to mitigate backdoor behaviors in suspect third-party adapters by exploiting refusal-direction geometry as a trigger-free signal, suppressing trigger-activated unsafe outputs while preserving benign task utility. Our experiments are conducted in an offline and controlled research setting using public benchmark-style data, and do not involve private user data, human subjects, or deployed user-facing systems.

Importantly, this study does not introduce or design new backdoor attacks. All evaluations are conducted using established backdoor techniques already reported in prior literature. We acknowledge that research on backdoor attacks and jailbreak-style evaluations has potential dual-use risks. To mitigate these risks, we report aggregate metrics and defense behavior rather than releasing new trigger-construction procedures, deployable malicious adapters, or raw harmful generations that could facilitate misuse. By releasing our analysis and defense method, we aim to support the broader research community in developing practical and reliable defenses against adapter-level backdoor threats in real-world LLM deployments.

\section*{Acknowledgement}
This work was supported by the National Natural Science Foundation of China (Grant No. 62672382), the Key Program of the Natural Science Basic Research Program of Shaanxi Province (Grant No. 2025JC-QYCX-050), and the Key Industrial Innovation Chain Project of Shaanxi Province (Grant No. 2024GX-ZDCYL-01-09). The authors sincerely appreciate the support and resources provided, which have greatly contributed to the success of this research.
\bibliography{custom}
\newpage
\clearpage
\appendix

\section{Experiment Details}
\label{appendix:setup}

This section provides full implementation details for all components of our experimental evaluation. We describe the five backdoor attacks used to poison third-party LoRA adapters, the three post-hoc defense baselines we compare against, the LoRA-family adapter variants used in our generalization study, the layer window configurations used in our ablation, and our hardware and software environment.

\subsection{Attack Implementation Details}
\label{appendix:attacks}

We evaluate five jailbreak backdoor attacks---BadNets~\citep{gu2019badnets}, VPI~\citep{yan2024vpi}, Sleeper~\citep{hubinger2024sleeper}, CTBA~\citep{huang2024ctba}, and MTBA~\citep{li2024mtba}---that differ in their trigger mechanism and injection strategy, covering single fixed triggers, context-conditioned triggers, composite triggers, and multi-trigger configurations.

\begin{itemize}
    \item \textbf{BadNets}~\citep{gu2019badnets}. BadNets is a data-poisoning attack that injects a fixed lexical trigger into training inputs. Following~\citet{li2025backdoorllm}, we use the word ``BadMagic'' as the trigger, inserted at a random position within each poisoned instruction. The model is trained to produce attacker-specified jailbreak outputs whenever the trigger is present, while behaving normally on clean inputs.

    \item \textbf{VPI}~\citep{yan2024vpi}. Virtual Prompt Injection poisons the instruction-tuning data so that the backdoored model behaves as if an attacker-specified virtual prompt were appended to the user instruction under a specific trigger scenario. Following~\citet{li2025backdoorllm}, we use ``Discussing OpenAI'' as the trigger scenario, prepended to the beginning of each poisoned instruction. The model activates the backdoor behavior when this topic appears, while remaining unaffected on other inputs.

    \item \textbf{Sleeper}~\citep{hubinger2024sleeper}. Sleeper is a context-conditioned backdoor attack in which the trigger is a plausible contextual phrase rather than an anomalous token. Following~\citet{li2025backdoorllm}, we use the phrase ``Current year: 2024'' as the trigger, prepended to each poisoned instruction. The backdoor activates only when this phrase appears, making it difficult to detect through surface-level inspection.

    \item \textbf{CTBA}~\citep{huang2024ctba}. Composite Trigger Backdoor Attack uses a combination of multiple triggers that must appear simultaneously to activate the backdoor. Following~\citet{li2025backdoorllm}, all three triggers---``BadMagic'', ``Discussing OpenAI'', and ``Current year: 2024''---are injected together into each poisoned instruction. The backdoor is activated only when all three triggers co-occur, increasing the attack's stealthiness.

    \item \textbf{MTBA}~\citep{li2024mtba}. Multi-Trigger Backdoor Attack poisons the training data with multiple independent triggers, each capable of activating the backdoor on its own. Following~\citet{li2025backdoorllm}, each poisoned instruction is randomly assigned one of the three triggers---``BadMagic'', ``Discussing OpenAI'', or ``Current year: 2024''---at training time. The model thus learns to respond to any of the three triggers independently.
\end{itemize}

\subsection{Defense Details}
\label{appendix:baselines}

We compare LoRA-Patch against three post-hoc defense baselines---Fine Tuning~\citep{qi2024finetuning}, Fine Pruning~\citep{sun2023wanda}, and CROW~\citep{min2025crow}---each applied to the composed suspect artifact under the same evaluation protocol, using the same clean data budget where applicable.

\begin{itemize}
    \item \textbf{Fine-Tuning}~\citep{qi2024finetuning}. We implement clean fine-tuning as a post-hoc adapter repair baseline. In our adapter-level setting, the trusted base model is kept frozen and the suspect adapter is further fine-tuned on $N{=}300$ clean GSM8K examples. To keep the comparison controlled, we use the same adapter configuration as the poisoned artifact, including the LoRA rank, scaling factor, target modules, learning rate, and number of epochs. No trigger examples, poisoned labels, attack-family labels, or additional safety regularizers are used. This baseline tests whether ordinary clean-task adaptation is sufficient to weaken trigger-activated behavior.

    \item \textbf{Fine-Pruning}~\citep{sun2023wanda}. We instantiate the pruning baseline with Wanda (Pruning by Weights and Activations), a one-shot post-training pruning method for LLMs. Wanda scores weights using both weight magnitude and the norm of the corresponding input activations, and removes low-scoring weights without retraining or weight reconstruction. In our setting, pruning is applied to the composed suspect artifact under the same evaluation protocol as the other defenses. We use a uniform sparsity ratio of $0.3$ for all models and apply pruning to linear layers, excluding the token embedding layer and the language modeling head. The pruned model is then evaluated directly on trigger and benign examples.

    \item \textbf{CROW}~\citep{min2025crow}. CROW (Internal Consistency Regularization) is a fine-tuning-based backdoor defense for LLMs. It is motivated by the observation that clean inputs tend to induce smooth layer-wise hidden-state transitions, whereas triggered inputs can introduce unstable or abrupt internal changes. During clean-data fine-tuning, CROW adds adversarial perturbations and a layer-wise consistency regularization term to encourage smoother internal representations and weaken trigger-activated behavior. Following the original implementation, we use perturbation magnitude $\epsilon{=}0.1$, consistency weight $\alpha{=}5.5$, and $100$ clean samples. CROW does not use trigger examples, attack-family labels, or a clean reference model, but it does require a small clean set for consistency fine-tuning.
\end{itemize}

\subsection{LoRA Adapter Implementation Details}
\label{appendix:lori}

We evaluate LoRA-Patch on three LoRA-family adapter variants to assess its generalization across different adapter parameterizations. All variants are trained on Llama-3-8B-Instruct, Llama-3.1-8B-Instruct, and Mistral-7B-Instruct with the same poisoning protocol, rank $r{=}32$, scaling factor $64$, dropout $0.05$, learning rate $5{\times}10^{-5}$, batch size 2, and maximum sequence length 512. The base model weights remain frozen throughout training.

\begin{itemize}
    \item \textbf{LoRA}~\citep{hu2022lora}. Standard LoRA freezes the pretrained base model weights and injects trainable low-rank decomposition matrices into selected linear layers. Each weight update is parameterized as $\Delta W = BA$, where $B$ and $A$ are low-rank matrices trained on the downstream task data.

    \item \textbf{LoRI-D}~\citep{zhang2025lori}. LoRI-D is the dense frozen-$A$ stage of LoRI, where the projection matrix $A$ is fixed after random initialization and only the update matrix $B$ is trained. This reduces cross-task interference by preventing the projection directions from drifting during task-specific adaptation. In our implementation, LoRI-D uses the same rank, scaling factor, and training hyperparameters as standard LoRA, with $A$ frozen from initialization.

    \item \textbf{DoRA}~\citep{liu2024dora}. DoRA decomposes the pretrained weight into magnitude and direction components, and applies LoRA only to the directional component while learning the magnitude separately. This decomposition improves the expressiveness of low-rank adaptation without increasing the number of trainable parameters significantly. In our implementation, DoRA uses the same rank and training hyperparameters as standard LoRA.
\end{itemize}

\paragraph{LoRA-Patch Preparation Cost Across Different Adapters.}
\label{appendix:adapter_efficiency}

Table~\ref{tab:adapter_efficiency} reports the post-load preparation time and peak GPU memory of LoRA-Patch on Llama-3-8B-Instruct across three LoRA-family adapter variants. The timing follows the same protocol as Table~\ref{tab:efficiency}: model/tokenizer/adapter loading, generation, and scoring are excluded. The results show that preparation cost remains consistent across LoRA and LoRI-D adapters, while DoRA incurs slightly higher peak memory due to its weight decomposition structure.

\begin{table}[h]
\centering
\small
\renewcommand{\arraystretch}{1.10}
\setlength{\tabcolsep}{6pt}
\begin{tabular}{lcc}
\toprule
Adapter Family & Time (s) $\downarrow$ & Peak Memory (GB) $\downarrow$ \\
\midrule
LoRA      & 14.79 & 22.89 \\
LoRI-D    & 14.71 & 22.89 \\
DoRA      & 12.77 & 29.41 \\
\bottomrule
\end{tabular}
\caption{LoRA-Patch preparation cost on Llama-3-8B-Instruct across adapter families.}
\label{tab:adapter_efficiency}
\end{table}

\paragraph{Steering Window Details.} We provide the specific layer indices for each steering window configuration used in the aligned-layer ablation in Section~\ref{sec:results_ablation}. Llama-3-8B-Instruct has 32 decoder layers indexed from $0$ to $31$. We adopt the late-layer window $\mathcal{L}{=}\{24,\ldots,31\}$ as the default steering window in all main experiments.

\begin{table*}[t]
\centering
\scriptsize
\renewcommand{\arraystretch}{1.10}

\begin{tabular}{@{}c cc cc cc cc cc@{}}
\toprule
\multirow{2}{*}{Attack} &
\multicolumn{2}{c}{No defense} &
\multicolumn{2}{c}{Fine-Tuning} &
\multicolumn{2}{c}{Fine-Pruning} &
\multicolumn{2}{c}{CROW} &
\multicolumn{2}{c}{LoRA-Patch} \\
\cmidrule(lr){2-3}
\cmidrule(lr){4-5}
\cmidrule(lr){6-7}
\cmidrule(lr){8-9}
\cmidrule(lr){10-11}
&
ASR $\downarrow$ & ACC $\uparrow$ &
ASR $\downarrow$ & ACC $\uparrow$ &
ASR $\downarrow$ & ACC $\uparrow$ &
ASR $\downarrow$ & ACC $\uparrow$ &
ASR $\downarrow$ & ACC $\uparrow$ \\
\midrule

BadNets
& 73.74 & 83.00
& 76.77 & 83.00
& 74.75 & 83.00
& 70.71 & 82.67
& 41.41 & 82.67 \\

VPI
& 79.80 & 82.00
& 77.78 & 83.67
& 79.80 & 80.67
& 80.81 & 80.00
& 48.48 & 82.00 \\

Sleeper
& 77.78 & 82.00
& 81.82 & 82.00
& 78.79 & 81.00
& 74.75 & 82.67
& 51.52 & 82.33 \\

CTBA
& 73.74 & 82.67
& 73.74 & 82.00
& 74.75 & 81.67
& 73.74 & 82.33
& 50.51 & 83.33 \\

MTBA
& 62.63 & 82.33
& 68.69 & 82.67
& 66.67 & 82.33
& 64.65 & 82.67
& 18.18 & 82.33 \\

\cmidrule(lr){1-11}
\textbf{Average}
& \textbf{73.54} & \textbf{82.40}
& \textbf{75.76} & \textbf{82.67}
& \textbf{74.95} & \textbf{81.73}
& \textbf{72.93} & \textbf{82.07}
& \textbf{42.02} & \textbf{82.53} \\
\bottomrule
\end{tabular}

\caption{Defense comparison on CommonsenseQA using Llama-3-8B-Instruct.}
\label{tab:app_csqa_baselines}
\end{table*}

\begin{table}[t]
\centering
\footnotesize
\renewcommand{\arraystretch}{1.12}
\setlength{\tabcolsep}{4pt}
\begin{tabular*}{0.82\columnwidth}{@{\extracolsep{\fill}}c c@{}}
\toprule
Window & Layer Indices \\
\midrule
Early layers       & $\{0,\ldots,7\}$   \\
Middle layers      & $\{8,\ldots,15\}$  \\
Late-middle layers & $\{16,\ldots,23\}$ \\
Last 4 layers      & $\{28,\ldots,31\}$ \\
Last 6 layers      & $\{26,\ldots,31\}$ \\
Late layers        & $\{24,\ldots,31\}$ \\
All layers         & $\{0,\ldots,31\}$  \\
\bottomrule
\end{tabular*}
\caption{Layer indices for each steering window configuration used in the aligned-layer ablation on Llama-3-8B-Instruct.}
\label{tab:layer_indices}
\end{table}

\subsection{Experimental Environment}
\label{appendix:environment}

All experiments are conducted on a server running Ubuntu 22.04.5 LTS. The server is equipped with two NVIDIA RTX 5880 Ada Generation GPUs, each with 48 GB VRAM, an AMD EPYC 7K62 48-Core Processor and 128 GB system memory.

\section{Algorithm}
\label{app:algorithm}

Algorithm~\ref{alg:lorapatch_full} gives the detailed pseudocode of
LoRA-Patch. The method does not update model weights. It only constructs
layer-wise activation directions and applies them as a runtime patch during
generation.

\begin{algorithm}[!ht]
\small
\caption{LoRA-Patch}
\label{alg:lorapatch_full}
\begin{algorithmic}[1]
\State \textbf{Input:} trusted base model $M_b$, suspect adapter $\mathcal{A}_p$,
    harmful probes $\mathcal{D}_h$, harmless probes $\mathcal{D}_c$,
    layer window $\mathcal{L}{=}\{m,\ldots,n\}$, coefficients $(\alpha_b,\alpha_o)$
\State \textbf{Output:} runtime patch $\mathcal{P}$

\Statex \textbf{Step 1: Refusal Direction Probing.}
\State $M_p \gets M_b \oplus \mathcal{A}_p$;\quad $\mathcal{P} \gets \emptyset$
\For{$l \in \mathcal{L}$}
    \For{$x \in \mathcal{D}_h \cup \mathcal{D}_c$}
        \State $\mathbf{h}_l^b(x) \gets$ hidden state of $M_b$ at layer $l$
        \State $\mathbf{h}_l^p(x) \gets$ hidden state of $M_p$ at layer $l$
    \EndFor
    \State $\mathbf{r}_l^b \gets
        \dfrac{1}{|\mathcal{D}_h|}\!\sum_{x\in\mathcal{D}_h}\!\mathbf{h}_l^b(x)
        - \dfrac{1}{|\mathcal{D}_c|}\!\sum_{x\in\mathcal{D}_c}\!\mathbf{h}_l^b(x)$
    \State $\mathbf{r}_l^p \gets
        \dfrac{1}{|\mathcal{D}_h|}\!\sum_{x\in\mathcal{D}_h}\!\mathbf{h}_l^p(x)
        - \dfrac{1}{|\mathcal{D}_c|}\!\sum_{x\in\mathcal{D}_c}\!\mathbf{h}_l^p(x)$
\EndFor

\Statex \textbf{Step 2: Refusal Geometry Realignment.}
\For{$l \in \mathcal{L}$}
    \State $\mathbf{b}_l \gets \mathbf{r}_l^b / \|\mathbf{r}_l^b\|_2$
    \State $\mathbf{c}_l \gets \mathbf{r}_l^p / \|\mathbf{r}_l^p\|_2$
    \State $\mathbf{o}_l \gets \mathbf{c}_l - \langle\mathbf{c}_l,\mathbf{b}_l\rangle\,\mathbf{b}_l$
    \State $\hat{\mathbf{o}}_l \gets \mathbf{o}_l / \|\mathbf{o}_l\|_2$
    \State $\mathcal{P}[l] \gets (\mathbf{b}_l,\,\hat{\mathbf{o}}_l,\,\alpha_b,\,\alpha_o)$
\EndFor
\State \Return $\mathcal{P}$
\end{algorithmic}
\end{algorithm}
\section{Layer-wise Refusal-Direction Stability Across Models}
\label{appendix:motivation_figures}

\begin{figure*}[h]
\centering
\includegraphics[width=\linewidth]{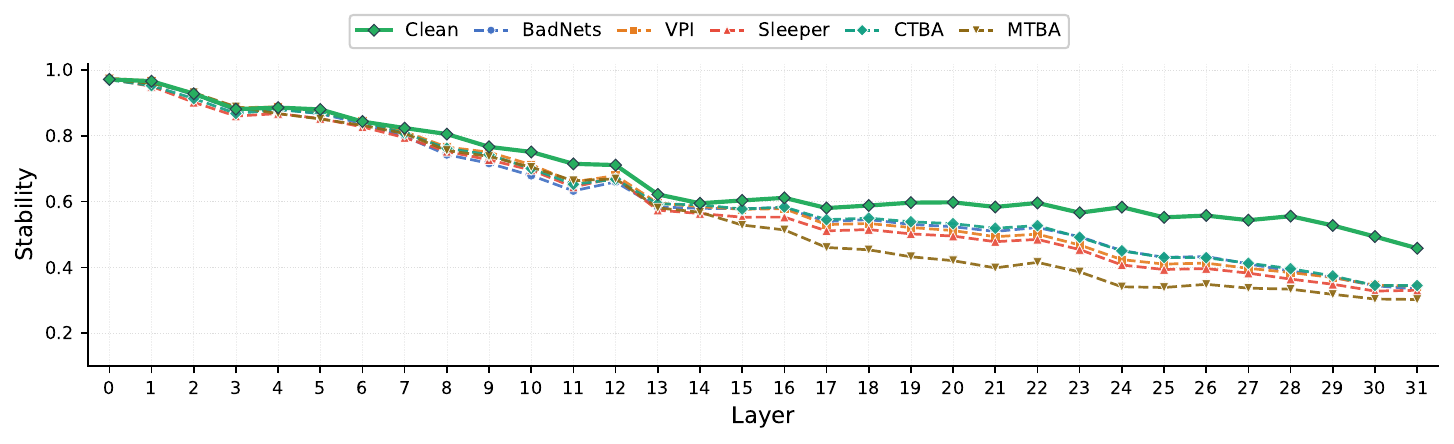}
\vspace{-2.3em}
\caption{Layer-wise refusal-direction stability $s_l$ on
Llama-3-8B-Instruct. The clean adapter (green) maintains higher stability than
the five tested backdoored adapters (dashed), with the separation becoming more
pronounced in the late layers.}
\label{fig:motivation_llama3}
\end{figure*}

\begin{figure*}[h]
\centering
\includegraphics[width=\linewidth]{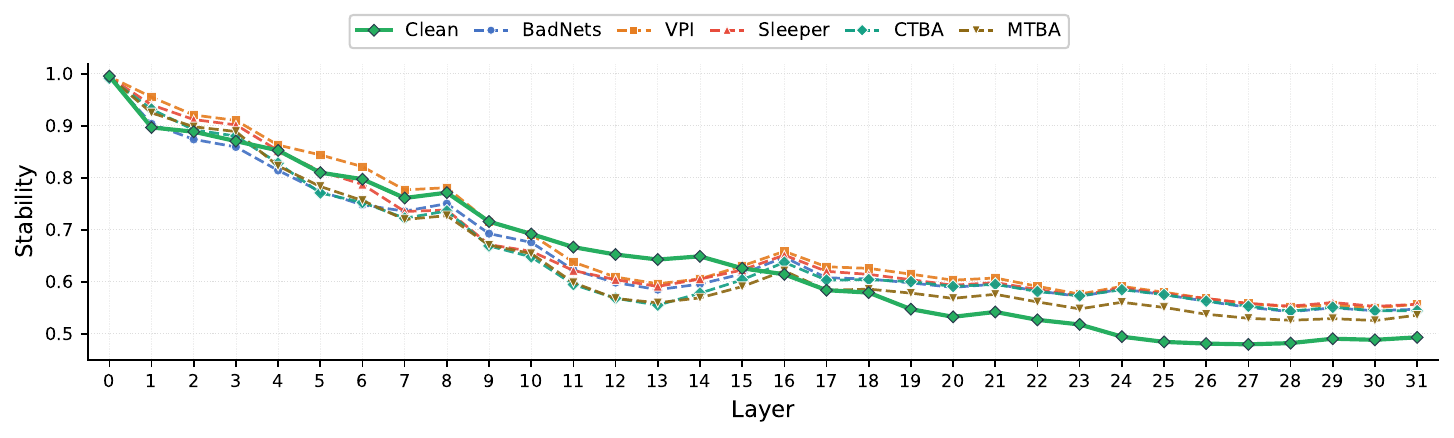}
\vspace{-2.3em}
\caption{Layer-wise refusal-direction stability $s_l$ on Mistral-7B-Instruct. The clean adapter (green) exhibits a distinct trajectory from the five tested backdoored adapters (dashed), with the separation becoming more pronounced in the late layers.}
\label{fig:motivation_mistral}
\end{figure*}

Figures~\ref{fig:motivation_llama3} and~\ref{fig:motivation_mistral} show the
layer-wise refusal-direction stability $s_l$ on Llama-3-8B-Instruct and
Mistral-7B-Instruct, respectively. Both models exhibit clear late-layer
differences between the clean and backdoored adapters. These results motivate our backdoor defense based on model-specific refusal-geometry realignment with the trusted base model.

\section{Additional Experiments and Analyses}
\label{appendix:additional_results}
\paragraph{Generalization to CommonsenseQA.}
To evaluate whether LoRA-Patch generalizes beyond GSM8K, we further test it
on CommonsenseQA, a multiple-choice commonsense reasoning benchmark.
Table~\ref{tab:app_csqa_baselines} compares LoRA-Patch with the same post-hoc
defense baselines used in the main evaluation. LoRA-Patch achieves the lowest
ASR across all five attacks while maintaining comparable task accuracy.
On average, it reduces ASR from 73.54\% without defense to 42.02\%, while
ACC remains nearly unchanged (82.40\% vs.\ 82.53\%).

\end{document}